\documentclass[mathleft, final
]{an}
\usepackage{graphicx}
\usepackage{times}
\begin{document}

\Pagespan{616}{}
\Yearpublication{2013}%
\Yearsubmission{2013}%
\Month{4}%
\Volume{334}%
\Issue{7}%
 \DOI{10.1002/asna.201311913}%

\title{An investigation into the radial velocity variability of GJ 581: \\[1.5pt]
         On the significance of GJ 581g}

\author{A.P. Hatzes\inst{}\thanks{Corresponding author:
  {artie@tls-tautenburg.de}}
}
\titlerunning{The radial velocity variations of the GJ 581 planetary system}
\authorrunning{A.P. Hatzes}
\institute{
Th\"uringer Landessternwarte, Sternwarte 5, D-07778 Tautenburg, Germany
}

\received{2013 Apr 15}
\accepted{2013 May 22}
\publonline{2013 Aug 1}


\abstract{%
We investigate  the radial velocity variations of GJ 581  based on measurements from
the HARPS and Keck HIRES spectrographs.
A Fourier pre-whitening procedure is able to extract four
planetary signals in the HARPS data and two from the Keck data. Combining both data sets increases the  significance of
the four planet signals found by HARPS. This indicates that the Keck data also supports the presence of
four planets. A periodogram analysis of the residual radial velocity measurements after removal of the
four planetary signals shows several periodic signals that are significant when assessing the false alarm 
probability via a bootstrap. However,
it is demonstrated that these are not due to planetary companions. This analysis is able to confirm the presence of
four planets around GJ 581, but not the presence of GJ 581g.  }

\maketitle

\section{Introduction}

Radial velocity (RV) searches for exoplanets has made great strides in improving
the precision of stellar RV measurements. Whereas 30 years ago Campbell \& Walker (1980)
achieved a breakthrough precision of 13 m\,s$^{-1}$, current RV surveys often achieve
a precision of $\approx$1 m\,s$^{-1}$, or a factor of 10 better. Consequently modern surveys
are finding exoplanets of lower and lower masses. 

Low mass planets often can produce a velocity $K$-amplitude comparable to the
measurement error or  intrinsic stellar variability due to activity or oscillations. 
Such planets are frequently found in multiple systems that only increase the difficulty in extracting the signal
of individual planets. This can lead to discovered planets that are refuted, or that remain controversial.
A case in point is the planetary system around the M dwarf GJ 581.  Bonfils et al. (2005) reported the presence of a 5.37-d hot
Neptune (GJ 581b), followed by the discovery of two additional planets with periods of 12.93~d, 
$m$ = 5.06 $\rm M_\oplus$ (GJ 581c) and $P = 83.4$ d (GJ~581d). Subsequently Mayor et al. (2009)
revised the period of GJ 581d to 67 d as well as provided evidence for a fourth planet with a period
of 3.15 d (GJ 581e). The evidence points to GJ 581 hosting  4 planets, although 
recent work by Baluev (2012) questions whether the fourth planet, GJ 581d is present.

 Vogt et al. (2010, hereafter V2010) presented 11 years of  RV  measurements
made  with the Keck HIRES spectrograph and combined these with the available HARPS
measurements. They  could confirm the presence of the four planets, but they 
also claimed
the presence of a fifth, GJ 581g, with a minimum mass
of 3.1  $\rm M_{\oplus}$  and a period of 36 d. This period placed GJ 581g firmly in the habitable
zone of the host star. The presence of this signal seemed to be significant, having a false
alarm probability of  $\approx$10$^{-7}$. Habitable planets are of considerable interest to the community 
and this discovery prompted work on both the habitability  of this fifth planet
(Heng \& Vogt 2011; Pierrehumbert 2011;
von Bloh et al. 2011), as well as dynamical stability of the
GJ 581 system as a whole (Tadeu et al. 2012).

The planet GJ 581g proved to be controversial. 
Tuomi (2011) presented a Baysian re-analysis of the radial velocities for GJ 581
and concluded that only four planets were present. Stronger evidence against a fifth planet came from 
Forveille et al. (2011)  who used 121 additional HARPS measurements for this
star and these along with the previous ones  showed no evidence for the presence of GJ 581g.  However, GJ 581g
would ``not go gentle into that good night". Vogt, Butler, \& Haghighipour (2012, hereafter VBH2012) re-analyzed the HARPS data modeling 
it with self consistent orbits and concluded that the HARPS data indeed showed evidence for a 2.2 $\rm M_\oplus$
but with an orbital period of 32 d. Thus a low mass planet could still be in  the habitable zone of the star, although
the false alarm probability for this detection was high, at about 4\,\%. More recently, Gregory (2012) applied a Baysian
analysis to the RV data for GJ 581 and concluded that the GJ 581g was not present in the data. 

The purported presence of GJ 581g demonstrates the need for other  analyses of RV measurements
in order to confirm the discovery of low mass planets. In this paper we use an independent analysis
of all available RV measurements for GJ 581 to assess the presence and significance of signals that could
be attributed to planets in the
system. The tools utilized are simple Fourier pre-whitening and periodogram analyses -- the basic
tools employed by all planet hunters. The purpose of this paper is not only to confirm previous planets found in the GJ 581 system
(and possibly new ones), but also to see if these simple tools can reach the same conclusions
as more sophisticated methods such as Bayesian analysis.
It is also important to explore limitations and possible pitfalls when using these standard
tools to search for periodic signals in RV data and to assess their statistical significance.

\begin{figure*}
\includegraphics[bb= 1cm 2cm 28cm 19.5cm, width=150mm,height=100mm]{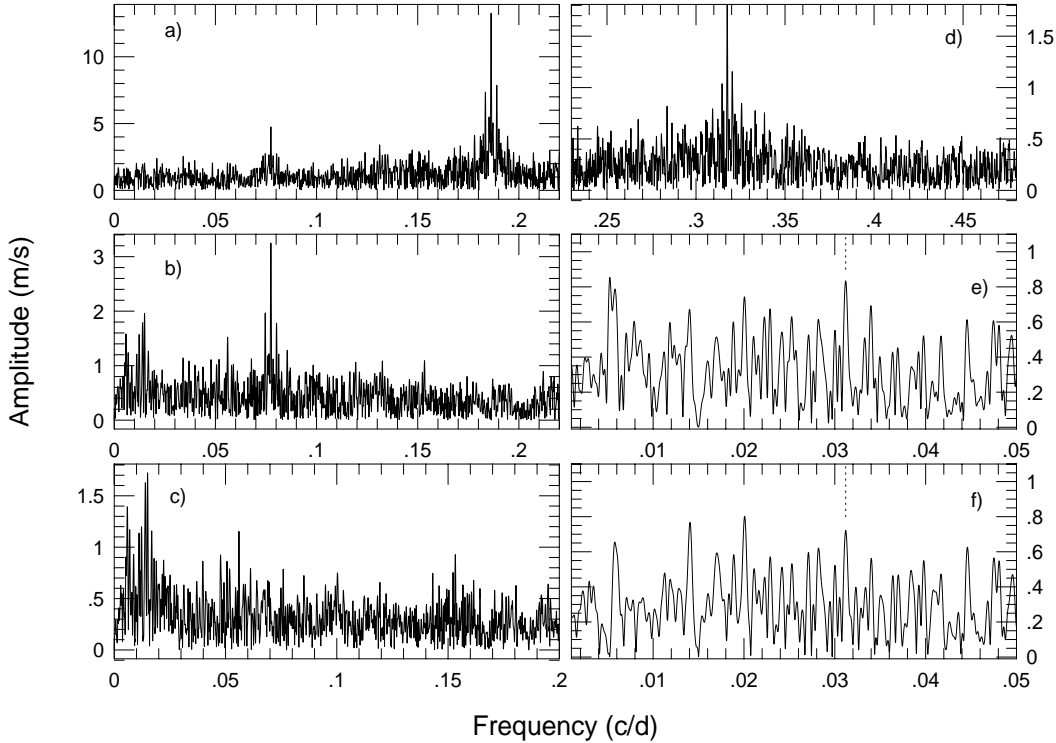}
\caption{Fourier pre-whitening performed on the HARPS RV data. \emph{Panel a}: amplitude spectrum
of the original RV measurements. \emph{Panel~b}: amplitude spectrum
after removing the sinusoidal variations from GJ 581b. \emph{Panel c}: amplitude spectrum after removing
both GJ 581b and c. \emph{Panel d}: amplitude spectrum after removing
GJ 581b, c, and d. \emph{Panel e}: amplitude spectrum after removing the highest
peak at $\nu = 0.0005$ c\,d$^{-1}$ from the previous panel. 
\emph{Panel f}: the final amplitude spectrum after removing the signals from the
four planets. The frequency shown by the vertical dashed line corresponds
to the 32-d period of VBH2012.
}
\label{harpspw}
\end{figure*}

\begin{figure*}
\includegraphics[bb= 1cm 2cm 28cm 19.5cm, width=150mm,height=100mm]{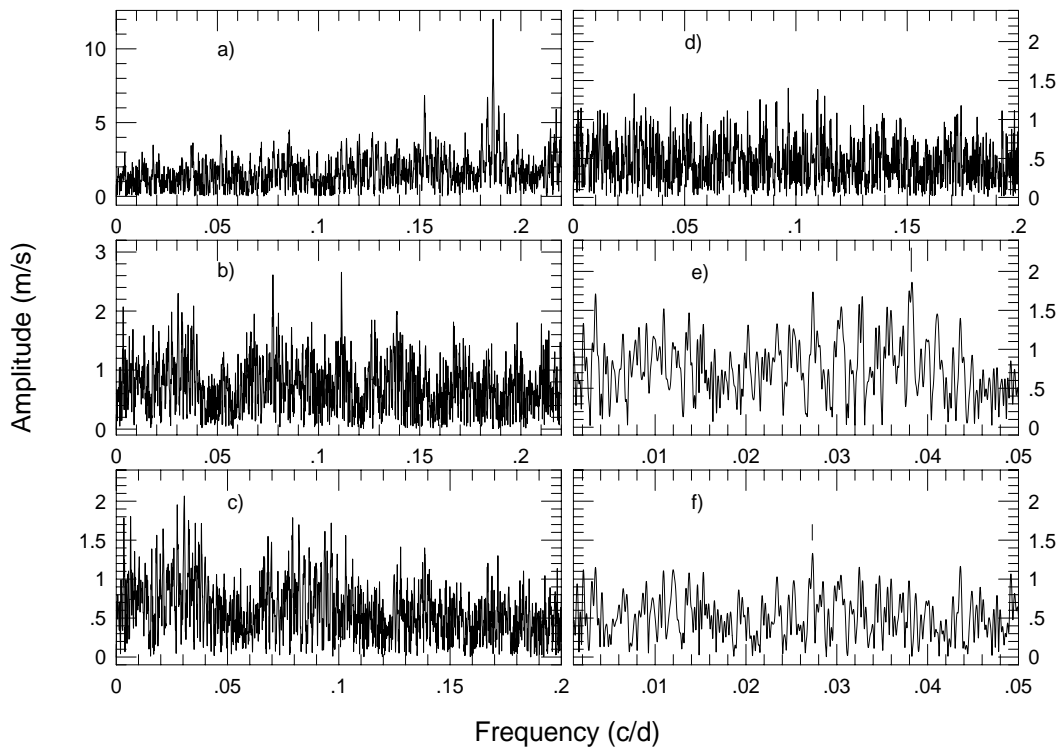}
\caption{Fourier pre-whitening performed on the Keck RV data. \emph{Panel a}: amplitude spectrum
of the original RV measurements. \emph{Panel~b}: amplitude spectrum
after removing the signal from GJ 581b. \emph{Panel c}: amplitude spectrum after removing
the alias frequency at 0.111 c\,d$^{-1}$. \emph{Panel d}: amplitude spectrum after removing
the true frequency to GJ 581c from Panel b.
\emph{Panel e}: same as previous panel, but for the frequency range 0.0 $<$ $\nu$ $<$
0.05 c\,d$^{-1}$. 
\emph{Panel f}: amplitude spectrum after removing the signal at 0.038 c\,d$^{-1}$ (shown by the vertical line in Panel e). The 
vertical line in Panel f marks the location of the 36-d period from V2010. }
\label{keckpw}
\end{figure*}

\begin{figure*}
\includegraphics[bb= 1cm 2cm 28cm 19.5cm, width=150mm,height=100mm]{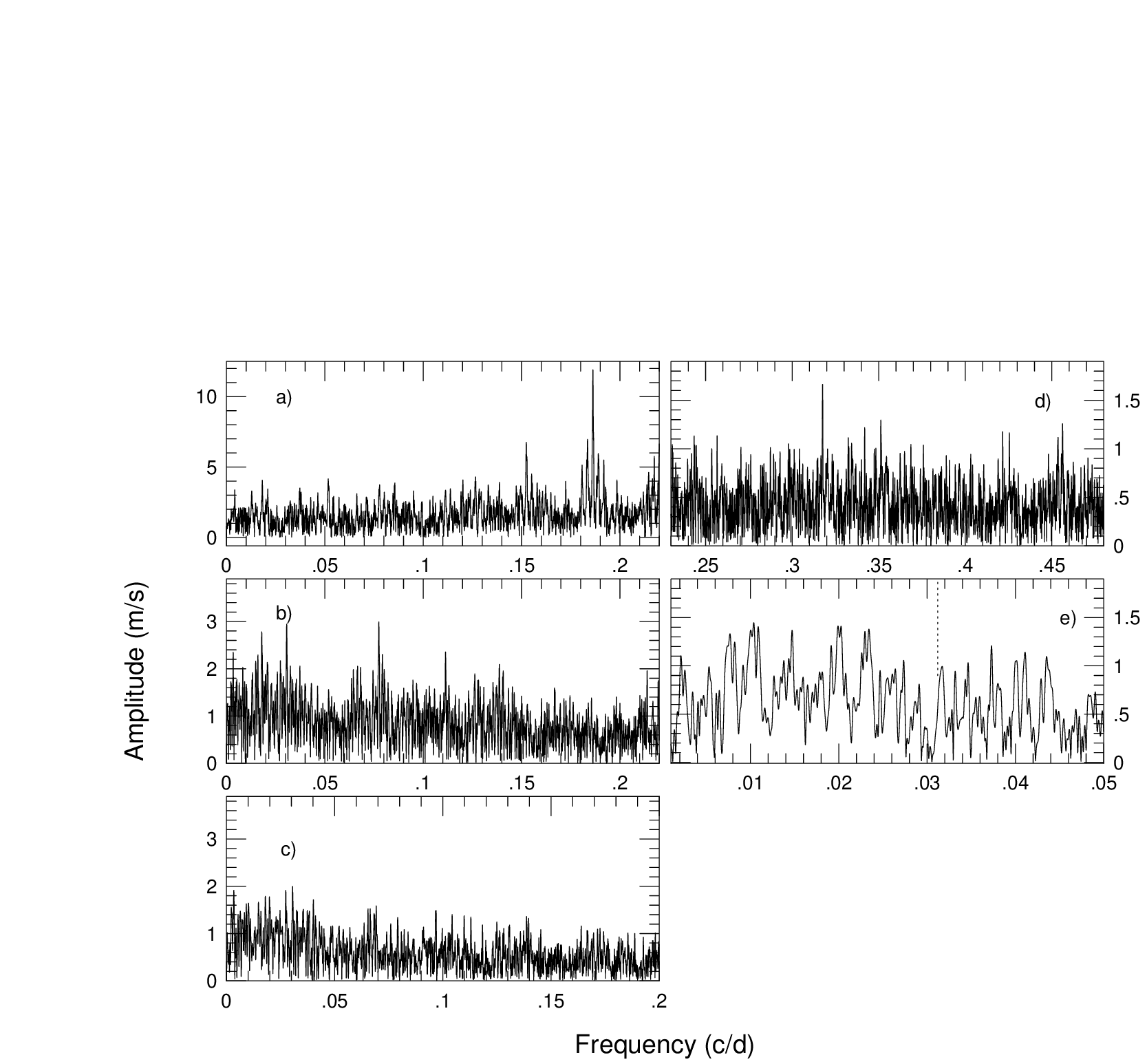}
\caption{Fourier pre-whitening performed on the Keck RV data but after first
removing the eccentric orbit of GJ 581d calculated from the HARPS data. \emph{Panel a}: amplitude spectrum
of the original RV measurements. \emph{Panel b}: amplitude spectrum
after removing the signal from GJ 581b. \emph{Panel c}: amplitude spectrum after removing
both GJ 581b and c. \emph{Panel d}: amplitude spectrum after removing
GJ 581b, c, and d. \emph{Panel e}: final amplitude spectrum after removing the
signals from the
four planets. The vertical dashed line 
marks the frequency  of the 32-d period signal found by VBH2012. }
\label{keckbpw}
\end{figure*}

\section{Frequency analysis with circular orbits}

\subsection{Combining the data sets}

The Keck and HARPS data sets have different zero points that must be applied to the
data before analyzing the combined data set. One can do this by calculating and subtracting
the mean of each data set, or by determining the offset of data taken on, or close to the
same nights. However, the presence of multi-periodic signals and measurement error may result in a
poor determination of the offset values. On the other hand, the presence of a known
periodic signal in the data provides a more robust way of determining this offset.
For this work we used the periodic variations of planet~b -- the dominant signal to the data,
We first analyzed the  HARPS data since it was the most extensive and was of slightly better quality than
the Keck data
(lower rms scatter). Since the presence of the other planet (GJ 581c, d, and e) could influence
the fitting of the orbit to planet b, the contribution of these
were removed from the RV measurements.

A Fourier pre-whitening procedure (see below) was used to fit sine functions (i.e. circular orbits)
to the orbital motions of GJ 581b,c,d,e as these were clearly in the data (see below). The
variations of the 3 planets c--e were then removed leaving behind the orbital motion of GJ 581b.
An orbital solution was then made to derive  the amplitude, period, and  ephemeris of planet b.

The pre-whitening procedure was then applied to the Keck data, but in this case only 
two planets were removed as supported by the pre-whitening process (see below). The result was two data sets 
containing the orbital variations predominantly from GJ 581b. 
 An orbital solution was then made to the two data sets keeping 
the amplitude, period, ephemeris of planet b fixed  to the 
values determined from the HARPS data, but  allowing the zero point offset to vary.
In this way a best fit, in a least squares sense, was made to the zero point offsets between the
two data sets. This offset was  applied to the original data with all planets, intrinsic variability, and noise
still present in the data.

\subsection{Fourier analysis with pre-whitening}

Fourier analysis with pre-whitening is a standard procedure in the field of stellar oscillations
for extracting multi-periodic signals from time series data. One finds the highest peak in the
Fourier amplitude spectrum, calculates a least squares sine fit (period, amplitude, and phase),
and subtracts this from the data. One then searches for additional signals often taken as the
highest amplitude in the residual amplitude spectrum.
This process continues until one reaches the level of the noise. This procedure should work
well on multi-planet systems in circular or low eccentricity orbits. For this process
the program {\it Period04} (Lenz \& Breger 2004) was used.

Tables 1--3
 show the results of this process for the HARPS data, Keck data, and combined data sets respectively.
In Table 1 the first four frequencies are due to planets, whereas the last two frequencies (separated by a line)
are of unknown origin.  In Table 2 the last frequency (below the line) is the alias frequency.
The rms scatter after removing the contributions of planets b--e is
1.98 m\,s$^{-1}$ for the HARPS data and 2.84 m\,s$^{-1}$ for the Keck data. 

\begin{table}
\caption{Periods found by pre-whitening of the  HARPS data.}
\label{tlab}
\begin{tabular}{ccc}\hline\noalign{\smallskip}
Frequency   & Period   &  $K$-amplitude \\ 
(d$^{-1}$)  &  (d)  & (m\,s$^{-1}$)  \\
\hline\noalign{\smallskip}
0.18627 $\pm$ 2.8$\times$10$^{-6}$ &  5.369  $\pm$ 8.4$\times$10$^{-5}$   & 12.49 $\pm$ 0.18\enspace  \\
0.07740 $\pm$ 1.2$\times$10$^{-5}$ &  12.920 $\pm$ 0.002                  & 3.29  $\pm$   0.12  \\
0.01499 $\pm$ 2.5$\times$10$^{-5}$ &  66.71  $\pm$ 0.01                   & 1.77  $\pm$ 0.22   \\
0.31754 $\pm$ 2.6$\times$10$^{-5}$ &  3.149  $\pm$ 2.5$\times$10$^{-5}$   & 1.81  $\pm$ 0.14  \\
\hline\noalign{\smallskip}
0.00519 $\pm$ 3.3$\times$10$^{-5}$ &  192.68 $\pm$ 1.22                   & 1.06 $\pm$ 0.17   \\
0.02008 $\pm$ 3.8$\times$10$^{-3}$ &  49.80  $\pm$ 0.09                   & 0.90 $\pm$ 0.24  \\
\hline
\end{tabular}
\end{table}

\begin{table}
\caption{Periods found by pre-whitening of the Keck data.}
\label{tlab}
\begin{tabular}{ccc}\hline\noalign{\smallskip}
Frequency   & Period   &  $K$-amplitude \\ 
(d$^{-1}$)  &  (d)  & (m\,s$^{-1}$)  \\
\hline\noalign{\smallskip}
0.18627 $\pm$ 4.4$\times$10$^{-6}$  & \enspace 5.370 $\pm$ 0.0001               & 12.35 $\pm$ 0.40\enspace \\
0.07744 $\pm$ 1.9$\times$10$^{-5}$  & 12.92  $\pm$ 0.003                &  2.93 $\pm$ 0.40 \\
0.03827 $\pm$ 2.6$\times$10$^{-6}$  & 26.17  $\pm$ 0.018                &  1.73 $\pm$ 0.40 \\
0.02734 $\pm$ 3.2$\times$10$^{-6}$  & 36.54  $\pm$ 0.042                &  1.43 $\pm$ 0.40 \\
\hline
\end{tabular}
\end{table}

\begin{table}
\centering
\caption{Periods found by pre-whitening of the combined data.}
\label{tlab}
\begin{tabular}{ccc}\hline\noalign{\smallskip}
Frequency   & Period   &  $K$-amplitude \\ 
(d$^{-1}$)  &  (d)  & (m\,s$^{-1}$)  \\
\hline\noalign{\smallskip}
0.18627 $\pm$ 3.8$\times$10$^{-6}$ &    5.3685 $\pm$ 0.0001 & 12.47 $\pm$ 0.21\enspace \\
0.07742 $\pm$ 1.5$\times$10$^{-5}$ &    12.9165 $\pm$ 0.0025\enspace &  3.29 $\pm$ 0.21 \\
0.01500 $\pm$ 1.4$\times$10$^{-5}$ &    66.667  $\pm$ 0.062\enspace  &  1.81 $\pm$ 0.22 \\
0.31756 $\pm$ 1.5$\times$10$^{-5}$ &    3.1496 $\pm$ 0.0001  &  1.83 $\pm$ 0.12 \\
\hline
\end{tabular}
\end{table}

Figures~\ref{harpspw}  and \ref{keckpw} show the results of the pre-whitening procedure applied to the HARPS and Keck data, respectively.
For clarity, narrow frequency windows about the planet signals are shown. In all figures of the
amplitude spectra or periodograms the unit of
choice is frequency (c\,d$^{-1}$) which is the natural unit of Fourier transforms and periodograms. Plotting the amplitude (power)
as a function of period results in a non-linear scale along the ordinate and
this distorts the periodogram. In the discussion both
frequency and period units will be used.
We only show the individual sets to highlight the differences in results between the two and because the
results for the combined data sets follow those using only the HARPS data. 

The Fourier analysis of only the
HARPS data was able to find all four planets with periods and amplitudes consistent with the combined data. The pre-whitening
of the HARPS data first finds planet b (Panel a in Fig.~\ref{harpspw}), then planet c (Panel b), followed by planet e (Panel c)
and finally Planet e (Panel d). Removing GJ 581b--e from the data results in
dominant peak at $\nu = 0.00519$ c\,$^{-1}$
($P = 192.6$~d).
However, there is second peak almost with the same amplitude at 0.0312 c\,$^{-1}$ which corresponds to the 32-d
period reported by VBH2012 (marked by the dashed vertical line). Removing the dominant peak at 192-d results in
approximately 4 equally strong peaks in the frequency range 0 $<$ $\nu$ $<$ 0.03 c\,d$^{-1}$, with the strongest
at 0.02 c\,$^{-1}$ ($P =49.8$ d). This and the 192-d period are listed in Table 1, but are not considered to be significant.
The 32-d period is  still present, but with reduced amplitude. 

The Fourier analysis of the Keck only data shows slightly different results, reflecting the few data points 
and slightly noisier
data. The pre-whitening finds planet b (Panel a in Fig.~\ref{keckpw}.). 
After removing planet b there are two strong peaks present in the amplitude spectrum: The
dominant peak was at a frequency of 0.111 c\,d$^{-1}$ and the true peak with a slightly lower amplitude
at 0.0774 c\,d$^{-1}$ ($P = 12.92$~d). The former is clearly an alias of the
true frequency as removing the true peak at 0.0774 c\,d$^{-1}$ also removes the
peak at $\nu = 0.111$ c\,d$^{-1}$. The pre-whitening procedure is 
unable to find either planet d or e. Table 2 shows the solution with the true frequency  for GJ 581c.

The residuals after removing the first two planets shows a peak near 26 d (${\nu
= 0.038}$ c\,d$^{-1}$)
(Panel e in Fig.~\ref{keckpw}). Removing this frequency shows a peak (Panel f)
at $\nu = 0.0273$ c\,d$^{-1}$ corresponding to the 36\,d period found by V2010.
Although we list these signals in Table 2 we do not consider these as significant.

\subsection{Analysis with Keplerian orbits}

The previous analysis employed multi-component sine fitting which is valid so long as 
the orbits are circular. Exoplanets can have significant eccentricity, even for 
short period planets. In Fourier space an eccentric orbit manifests itself by the 
presence of harmonics to the orbital frequency (2$\nu_{\rm orbit}$, 3$\nu_{\rm orbit}$, etc.).
In particular the 32-d period seen in the HARPS RV residuals is close to the 
first harmonic (2$\nu_{\rm orbit}$) of GJ 581d and might be due to an eccentric orbit.

 To investigate the presence of eccentric orbits a multi-component sine fit was made for the
 4-planet solution. The orbits of three planets were then subtracted leaving the orbital motion of 
 the remaining planet. An orbital solution was then made to this planet allowing the eccentricity
 to vary. This was successively done for all four planets in the GJ~581 system.

 \begin{table}
\caption{Eccentric orbital solution for GJ 581d.}
\label{tlab}
\begin{tabular}{ll}
\hline\noalign{\smallskip}
Parameter   & Value  \\ 
\hline\noalign{\smallskip}
Period (d) & 66.64  $\pm$ 0.08 \\
$K$ (m\,s$^{-1}$)    &  2.20 $\pm$ 0.19  \\
$e$     & 0.205 $\pm$ 0.08 \\
$\omega$ (\degr) & 2  $\pm$ 23 \\
$T_0$ & 245141.6407 $\pm$ 4.19 \\
\hline
\end{tabular}
\end{table}

\begin{table}
\caption{Fourier pre-whitening of the combined data after removal of eccentric
orbital solution for GJ 581d.}
\label{tlab}
\begin{tabular}{cc}\hline\noalign{\smallskip}
Period  & $K$-amplitude \\ 
(d) &  (m\,s$^{-1}$ )   \\
\hline\noalign{\smallskip}
5.3706 $\pm$ 0.0001 & 12.49 $\pm$ 0.17\enspace  \\
12.920 $\pm$ 0.002\enspace & 3.30  $\pm$   0.17  \\
3.150  $\pm$ 0.001 & 1.87  $\pm$ 0.17   \\
\hline
\end{tabular}
\end{table}

\begin{table}
\caption{Pre-whitening of Keck data with an eccentric GJ 581d.} 
\label{tlab}
\begin{tabular}{cc}\hline\noalign{\smallskip}
Period  & $K$-amplitude  \\ 
(d) &  (m\,s$^{-1}$ )   \\
\hline\noalign{\smallskip}
5.3710 $\pm$ 0.00001 & 12.47 $\pm$ 0.32\enspace \\
12.917 $\pm$ 0.001\enspace\enspace  & 3.30  $\pm$  0.32 \\
3.149  $\pm$ 0.0002 & 1.76  $\pm$ 0.32  \\
\hline
\end{tabular}
\end{table}
 
 The results of this analysis
 showed that the orbital motion of GJ 581b, c, and e were consistent with circular orbits.
 Either the Keplerian fit produced a slight eccentricity that was less than the error,
 or an unphysical slightly negative eccentricity, again to within the error of being zero. 
Circular orbits (sine functions) seem to be a 
 valid assumption for the orbital motion of GJ 581b, c, and e.
 Only GJ 581d showed significant eccentricity. Table 4 shows the final
 orbital solution for GJ 581d. The eccentricity of $e = 0.20 \pm$ 0.02 is
 consistent with that found by Forveille et al. (2011).
 
 The pre-whitening analysis was repeated on all the data after removing the eccentric Keplerian
 motion of GJ 581d. The results are shown in Table 5 for the combined data sets. 
 The results are consistent with the analysis assuming circular orbits for all planets. Even for GJ 581d 
the Fourier analysis (i.e. circular orbits)
 results in a comparable $K$-amplitude,  1.74 $\pm$ 0.17 m\,s$^{-1}$  versus 2.20 $\pm$ 0.19 
 m\,s$^{-1}$

The pre-whitening of the Keck data after removing an eccentric orbit for planet d showed
a slightly improved result than the one assuming a circular orbit. This is shown in Fig.~\ref{keckbpw}.
Removing an eccentric orbit for planet d suppresses the alias frequency at $\nu
= 0.111$ c\,d$^{-1}$
(Panel b)  and causes the true frequency at $\nu = 0.077$ c\,d$^{-1}$ to be the highest peak.
The residuals after subtracting planet c results in a peak at $\nu = 0.030$ c\,d$^{-1}$ of unknown origin, but certainly
not planetary in nature. Eliminating this signal results in the clear detection of planet e at 0.317 c\,d$^{-1}$ (Panel d). The results
of this pre-whitening procedure is shown in Table 6. The periods and amplitudes of GJ 581b,c, and e are
recovered with the correct periods and amplitudes. This provides some additional support
that GJ 581d has modest eccentricity. The remaining frequency in the data is still at 
$\nu = 0.027$ c\,d$^{-1}$, but with a reduced amplitude.

\subsection{Signals from a ``fifth'' planet}
We investigated the presence of periodic signals from a possible 
fifth planet in the system. A Scargle periodogram was employed since the power of a signal is
related to the statistical significance of the signal. As a personal rule of thumb, Scargle power, $z$, greater
than 10 is a sign of a possibly significant signal in the data.  Figure~\ref{ftall1}  shows the periodogram of the
RV residuals after subtracting circular orbits (i.e. pre-whitening) calculated using the combined data sets.
The panels shows the periodogram of the residuals of the Keck, HARPS, and combined data separately.
The Keck data shows a significant peak at $\nu = 0.0273$ c\,d$^{-1}$, the 36-d period reported by V2010.

The HARPS data do show evidence for other, possibly significant peaks. There is a peak at 
$\nu = 0.00523$ c\,d$^{-1}$ ($P = 192$ d) and another at 
$\nu = 0.0058$ c\,d$^{-1}$ ($P = 172$ d). There is also an equally strong peak
at $\nu = 0.031$ c\,d$^{-1}$. This corresponds to the 32-d period claimed by VBH2012. However, this peak is diminished
in the combined data set as are  the ones near $\nu = 0.00523$ c\,d$^{-1}$. Although the frequency is near
the harmonic of  the 67-d period, an eccentric orbit can not entirely reproduce the observed amplitude.
Tests made with a synthetic eccentric orbit with a period of 67 d and a $K$-amplitude of 
2 m\,s$^{-1}$ resulted in an amplitude  the harmonic of no more than 0.5 m\,s$^{-1}$, even for highly
eccentric orbits. This is significantly less than the amplitude of 0.9
m\,s$^{-1}$ seen at $\nu = 0.03$ c\,d$^{-1}$,
 The only remaining possible significant peak in
the combined data set is at $\nu = 0.02292$ c\,d$^{-1}$ ($P = 43.6$ d).

Figure~\ref{ftall2} shows the periodograms after removing an eccentric orbit for GJ 581d, and circular orbits for GJ 581b,c, and e.
They are similar, but with notable differences. In the Keck data the peak at
$\nu = 0.0273$ c\,d$^{-1}$ (36 d)
is greatly reduced (less significant) 
 In the HARPS data the peak at $\nu = 0.0052$ c\,d$^{-1}$ ($P = 192$ d)
 clearly becomes the dominant peak which vanishes in the combined data (lower panel Figure~\ref{ftall2}). 

 \section{Statistical significance of frequencies}

Here we investigate the statistical significance of the planet signals. For  a  Scargle periodogram the
power, $z$,  is related to the statistical significance by FAP = $1 - (1-{\rm e}^{-z})^N$ $\approx$ $N\,{\rm e}^{-z}$, where $N$ is the number of
independent frequencies which can be approximated by the number of data points. If the signal is incoherent, or short-lived,
adding more data would cause the Scargle power to decrease.  On the other hand for a  real signal,
more data would add to the significance of the detection, thus increasing the Scargle power.
Scargle periodograms  were performed on data containing a 
single ``planet'' signal after removing the
contribution of the other planets. 

\begin{figure}
\vskip-2.5cm
\hskip-3mm
\includegraphics[width=83mm,height=80mm]{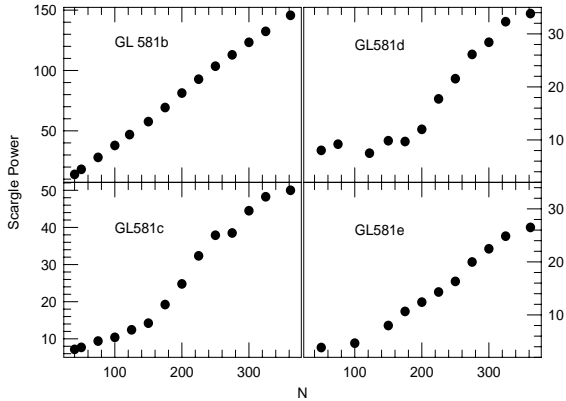}
\caption{The Scargle power as a function of the number of data points used for the periodogram. 
Counterclockwise from \emph{top left}: GJ 581b,c,d, and e. For each panel the contributions of the other planets
were removed before calculating the periodogram. }
\label{fapsall}
\end{figure}

Figure~\ref{fapsall} shows the power in the Scargle periodogram at the planet frequency
as a function of the number of data points for GJ 581b, c, d, and e.  First the HARPS data was
used, followed by adding the Keck data.
All four planets show
the expected behavior for  a real signal that is coherent and long lived -- the power and thus
significance of the signal increases with more data. Even though the Keck 
data showed evidence for only two planets,
the fact that the overall Scargle power for all planets
increases when adding these to the HARPS data  
 supports the fact that
GJ 581b-e are indeed all present in the Keck data.

\begin{figure}
\vskip-2.5cm
\hskip-3mm
\includegraphics[width=83mm,height=80mm]{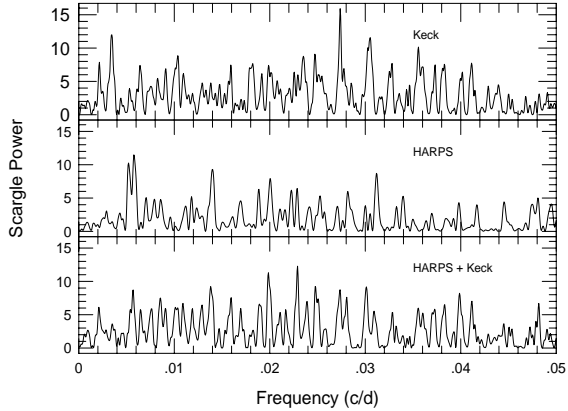}
\caption{\emph{Top}: the Scargle periodogram of the Keck data after performing a pre-whitening procedure
on the full data set (Keck + HARPS) and removing the signals of the four planets. \emph{Middle}:
 the same  as for the top panel but for the 
HARPS data. \emph{Bottom}: the Scargle period of the residuals (all planets removed)
for the entire data set. }
\label{ftall1}
\end{figure}

We now turn to the statistical significance of the  fifth planet, starting with residuals calculated
using circular orbits for all planets. The Keck RV residuals produced assuming circular orbits
for all planets show what appears to be a significant signal at ${\nu = 0.027}$ c\,d$^{-1}$.
The false alarm probability (FAP) of this signal  was assessed using a bootstrap randomization
procedure. The RV values were randomly shuffled 200\,000 times keeping the times fixed, performing periodograms for the
random data and then seeing how many random periodograms had power greater than the 
observed periodogram. The FAP determined in this way was 3.5$\times$10$^{-5}$.
The periodogram of the residual HARPS data (central panel Fig.~\ref{ftall1}), on the other hand, 
shows no evidence for the 36-d period as noted by Forveille et al. (2012). 

\begin{figure}
\vskip-2.5cm
\hskip-3mm
\includegraphics[width=83mm,height=80mm]{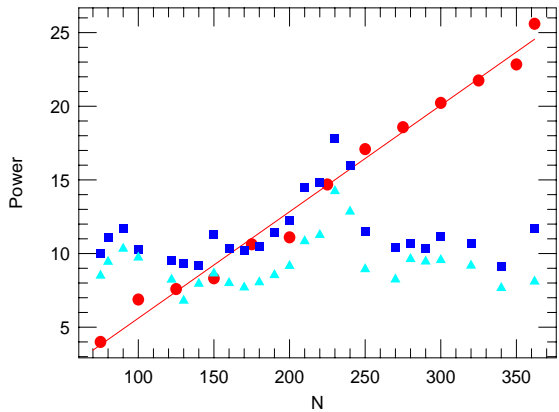}
\caption{Squares: the behavior of the Scargle power of the 36-d period found in the residual
Keck data. The data is in chronological order so Keck and HARPS data are mixed. At $N$ = 230
there are 100 Keck measurements and 130 HARPS measurements. Dots: the behavior of a fake signal with the
a period of 36-d and amplitude of 1.3  m\,s$^{-1}$ sampled in the same manner as the real data. Random noise
at a level of $\sigma$ = 2 m\,s$^{-1}$. The line represents a least squares fit to the simulated data in order to 
guide the eye. The triangles represent is the same analysis, but removing an
eccentric orbit for GJ 581d. }
\label{fap36}
\end{figure}

However, the picture gets more complicated
when one examines only a portion of the HARPS data. The residual RV data from HARPS and Keck
were combined and sorted according to time. The periodogram was calculated using the first
50 points and subsequently adding more data. Figure~\ref{fap36} shows the Scargle power at the 36-d
period as a function of $N$, the number of data points. One can see an increase in the power (and thus
significance) up to ${N = 230}$. This point represents 100 Keck measurements and 130 HARPS measurements.
This behavior also seems to be present even when using an eccentric orbit for GJ 581d (triangles), although the overall
power is less than for the circular orbit case.

\begin{figure}
\vskip-2.5cm
\hskip-3mm
\includegraphics[width=83mm,height=80mm]{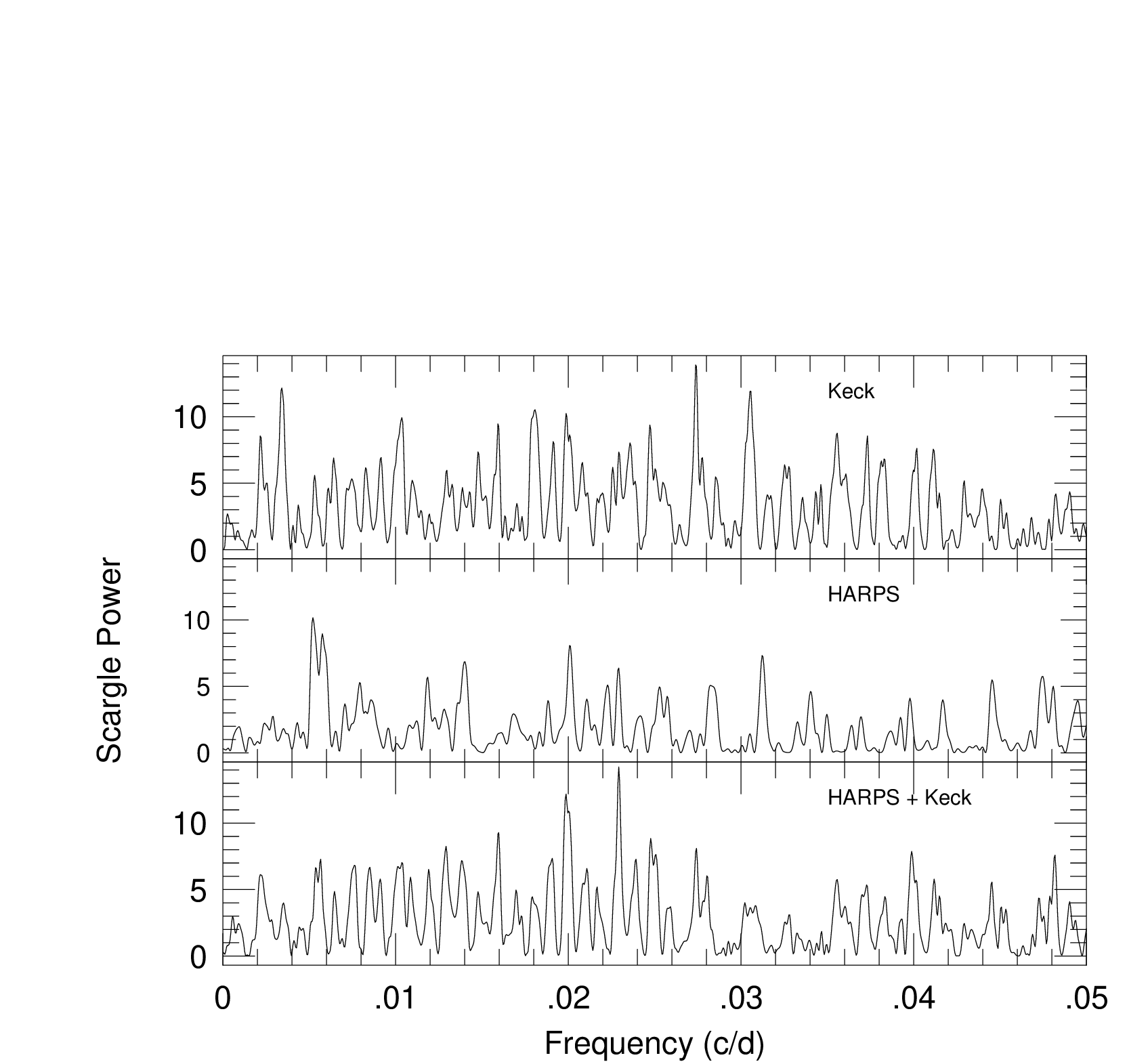}
\caption{The same as for Fig.~\ref{ftall1} but using an eccentric orbit for GJ
581d. }
\label{ftall2}
\end{figure}

To compare these results to those expected from a known real signal,  a sine function having a 36-d period and the same amplitude as the RV data was generated
and sampled in the same way as the data. Random noise with $\sigma$ = 2 m\,s$^{-1}$ was added to 
the simulated data. This value is comparable to the rms scatter of the HARPS data and only slightly less
than that of the Keck data ($\sigma$ = 2.7 m\,s$^{-1}$)
after removing the 36-d variation The dots in Fig.~\ref{fap36} represent the behavior of the Scargle power for these simulated data
and the line is a least square fit. The  power of the 36-d period
for the first 230 data points follows that of the fake data and  attains a maximum 
Scargle power of $z$ = 16  for the 36-d period. The FAP was determined using 200 000 random shuffles of the
data. In no instance did the power of the random data exceed the actual data periodogram and this
indicates a FAP $<$ 2$\times$10$^{-6}$. At face value this seems to indicate that the signal at 36-d in the Keck data is not 
only significant, but it is supported by part of the HARPS data.  Adding the first half of the  HARPS points indeed boosts
the significance of this signal causing the FAP to drop by a factor of 10.

The addition of the last 100 HARPS measurements, however, quickly causes the 36-d period
simply to go away. Including all the HARPS measurements causes the FAP of the 36-d period to drop
to $\approx$\,6\,\%. This signal if real is clearly not coherent or long-lived.

Table 7 lists all the possibly significant  periods (frequencies) and their amplitudes  found in the residual data after removing  
GJ 581b--d. Included is the 32-d period found by VBH2012. Listed are the data sets used
(Keck, HARPS, or combined) as well as the method:  Multi-component sine fitting which is the equivalent
to circular orbits (``Sine fitting''), or using an eccentric orbit for planet d, but circular orbits for the other 3 planets
(labeled ``orbit$_{\rm d}$ + Sine''). The FAP or these signals were estimated using a bootstrap of 200\,000 random shuffles
of the data. Note that most of these, with the exception of the 192-d period found in the HARPS only data, have an 
extremely low FAP. Virtually every  investigator (the author included) using RV data to search for planets would claim
these as significant detections of planetary signals.

\begin{figure}
\vskip-2.5cm
\hskip-3mm
\includegraphics[width=83mm,height=80mm]{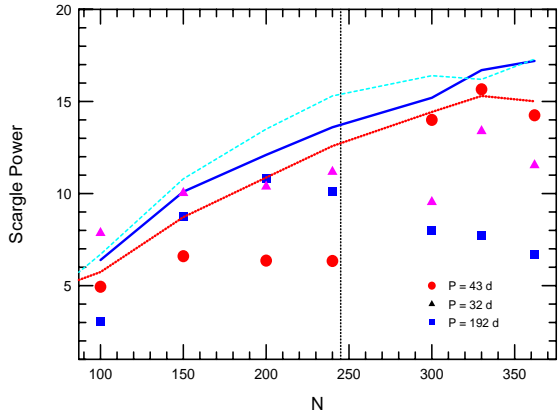}
\caption{The Scargle power of various periods (dots: 43 d, squares: 192 d,
triangles: 32 d) as a function of the number of data points used in the periodogram,
starting with HARPS data and then Keck data. The vertical dashed line marks the
boundary between the HARPS and Keck data (Keck data added to the right). The lines represent the results of simulated
data for the 32-d period (solid blue), the 43-d period (dotted red), and the
192-d period (dashed light blue).  }
\label{signals}
\end{figure}

To investigate the reality of these we can apply the
``significance versus number of points'' test. This was done by looking at the Scargle power of a signal by first
adding in succession the HARPS data, followed by the Keck data. This was done for 3 
periods: 32 d, 43 d, and 192 d and the results are shown in Fig.~\ref{signals}.  (The 176.7-d period behaves the same as the 192-d period and is not
shown.) The vertical dashed line marks the point at which   Keck data was added to the periodogram.  As a gauge of what one should expect for a real
signal, we generated a sine function with the same amplitude found in the analysis (Table 7) that was 
sampled in the same way as the real data. For the noise level we used 2 m\,s$^{-1}$ for the HARPS data and 2.7 m\,s$^{-1}$ for the Keck data, values consistent with the  actual ones.
The lines show the behavior of the power of a signal as a function of data points for the simulated data. It behaves as one
expects for a real signal - the power (significance) monotonically increases with more data.

None of the found periods show the behavior expected for a real signal. The 32-d and 192-d period show an initial increase
in significance, but that flattens out even in the HARPS data. Adding the Keck data shows a further drop in the significance
in the signal. The 43-d period does show an increase in power when adding the Keck data as does the 32-d period 
when adding the last few points. However, this behavior is highly suspect as one sees a dramatic increase only after adding
the Keck data. For example the 43-d period shows a increase in a factor of two in power
by adding the first 50 Keck points,
whereas adding the last 50 HARPS point to the data keeps the power constant. Most likely these sharp increases in the
Scargle power are related to different systematic errors in the data sets that cause false periods to appear due to
merging the data. In spite of the formal low FAP listed in Table 7, none of the periods listed there are deemed significant.

\begin{table}
\caption{Periods found in the residual RV data after removing 4 planets.}
\label{tlab}
\begin{tabular}{llrcc}\hline\noalign{\smallskip}
Set  &~~ Method & Period & $K$ &  FAP  \\ 
& & (d)~~~  &  (m\,s$^{-1}$ ) &  \\
\hline\noalign{\smallskip}
Keck & Sine fitting & 36.54 & 2.02 & 3.5$\times$10$^{-5}$  \\
HARPS & Sine fitting & 192.3 & 0.94 & 8.7$\times$10$^{-4}$ \\
All     &  Sine fitting  & 176.67   & 0.94  & 1.9$\times$10$^{-4}$ \\
Keck & orbit$_{\rm d}$ + Sine & 36.57 & 1.97 & 1.5$\times$10$^{-4}$ \\
HARPS & orbit$_{\rm d}$ + Sine  & 191.93  & 0.89 & 0.018 \\
All & orbit$_{\rm d}$ + Sine & 43.63 & 0.93 & 3.85$\times$10$^{-4}$\enspace \\
\hline
\end{tabular}
\end{table}

\newpage

\section{Discussion}
The results of this analysis of the RV measurements for GJ~581 confirms the presence of four planets
(GJ 581 b, c, d, and e) previously reported in the literature.  A reliable detection of four planets could be made with the 
HARPS data alone.  The statistical significance of these four planets
increases with more data and their  presence is also supported by independent
data from Keck HIRES. On the other hand, the other planetary signals reported in the literature, namely the
36-d and 32-d periods are not confirmed. GJ 581 is at the present time only a four planet system.

V2010 also reported a 
433-d period in their RV data.
Gregory (2012) in his analysis of the same data claimed that a period
of 399 d was present in the data which may be associated with the 433-d period found by V2010. 
The pre-whitening procedure could not find any significant peak at $\nu =
0.0025$ c\,d$^{-1}$ 
with an upper limit on the amplitude of 0.5 m\,s$^{-1}$. 
Thus, we find no strong evidence for a period of $\sim$ 400 days in the data.

The most important aspect of this analysis is not the confirmation of only four planets around GJ 581, but rather how one
can be fooled by apparently significant signals in the data that
are not real. For complicated multi-planet systems and especially with 
different data sets having different systematic errors,
sampling pathologies, etc., the answer you get depends on the details of the analysis. Spurious periodic signals can creep into 
frequency space when combining data sets with different systematic errors.  Unknown intrinsic stellar variability
that can be stochastic, incoherent, short lived, etc. coupled with the data window can also cause spurious periodic signals
that appear to be highly significant, only to disappear with more data. There is a lesson to be learned here: When dealing with complicated
RV variations and different data sets even highly significant signals should be treated with caution, and their nature
not so readily attributed to planetary companions. 

A case in point is the 36-d period first reported by V2010. Given the data available to the authors at the time of publication
a 36-d period was indeed present at a significant level and possibly due to another planet. Furthermore, the presence of this signal,
as shown here, was supported by two independent data sets. The authors reached a reasonable conclusion
based on the available data at hand. One can speculate whether this signal was in fact real, and possibly due to 
stellar variability. Interestingly, in their study of the M dwarf Barnard's star (GJ 699) K\"urster et al. (2003) also found RV variations with a
period of $\approx$ 32 d that appeared to be significant, but that showed correlations with variations of H$\alpha$ hydrogen
Balmer line  that was
interpreted as due to convective redshift. Possibly such similar investigations of H$\alpha$ may help to distinguish between variations
caused by planets and those due to stellar variability, particularly for M dwarf stars like GJ 581.

The 32-d period reported by VBH2012 is not confirmed by our study. In the periodogram a peak is seen in the HARPS
data at 32-d, but higher peaks are seen at other periods. Furthermore, the presence of this 32-d period should have become
more significant when adding the Keck data and it did not.  The behavior of the 32-d period when including more data is not
consistent with the presence of  coherent signal. To complicate matters, in the combined data sets the strongest peak in the periodogram
of the residual RV data is at 43 d, not 32 d, but this signal is not considered to be significant.

Low mass planets in the habitable zone of stars are of particular interest to researchers wishing to model their atmospheres and
assess the possibility of habitability.  The detection of these, however, will be challenging.
Many of these will be found in multiple systems and the
velocity reflex motion from a single planet may be comparable to the measurement error. 
Intrinsic stellar
variability from activity and systematic errors with an unknown frequency  spectrum may dominate the
variability of the bona fide planets. Furthermore, because of sparse sampling at one site we may be forced
to combine measurements from different sites, different instruments, and each with its own systematic
error characteristics. The amplitude spectra of the combined data may be complicated, and not just from
planetary signals.
However, planet hunters searching for such planets need to deliver the best possible candidates.
Investigators should not be investing time modeling atmospheres of non-existent planets.
To help confirm such planets there a several things to consider:

\begin{itemize}
\item It is essential to perform different analyses and with different approaches to see how 
robust periodic signals are in the data. GJ 581 is a case in point. Different groups analyzing the
same data set come to completely different conclusions using reasonable and valid
analysis tools. Such exoplanet discoveries should be treated
with caution.

\item When combining different data sets it is essential that detected periodic signals are present at some level in all
data sets. If a signal is not seen in one data set one should check that the statistical significance increases in an understandable way
when adding other data sets. Periods that suddenly appear when another
 data set is added should be treated with caution.

\item A low false alarm probability, even with FAP $<$ 10$^{-4}$,  for a periodic signal in a multi-planet system is no guarantee
that this signal is real. This is especially true if the amplitude of the signal is comparable to the measurement error.

\end{itemize}

Finally, this study demonstrates that traditional analysis tools like Fourier pre-whitening and Scargle periodograms are powerful
tools for extracting multi-periodic (i.e. planets) signals from RV data. These tools even work 
well in the case of eccentric orbits, one needs only to 
cautious about the interpretation of the harmonics of a main frequency. 
The Scargle periodogram is a standard and valuable
tool for assessing the FAP, but the ``user should beware''. Even highly significant detections with formally low FAP should be
treated with caution. It can be that in the process of removing other signals and thus variations in data, that this could artificially boost the
significance of a remaining signal. Examinations of subsets of the data as well as the use of simulated data are
needed to assess how significant a signal is.


\newpage


\begin{thebibliography}{}
  \bibitem{2013MNRAS.429.2052B} Baluev, R. V. 2012, MNRAS, 429, 2052
  \bibitem{2005A&A...443L..15B} Bonfils, X., Forveille, T., Delfosse, X.,  et al. 2005, A\&A, 443, L15 
  \bibitem{1979PASP...91..540C} Campbell, B., Walker, G.A.H.  1979, PASP, 91, 540
  \bibitem{2011arXiv1109.2505F} Forveille, T., Bonfils, X., Delfosse, X., et al.
   2011, astro-ph/1109.2505
  \bibitem{2011MNRAS.415.2523G} Gregory, P.C. 2012, MNRAS, 415, 2523 
  \bibitem{2011MNRAS.415.2145H} Heng, K., Vogt, S.S. 2011, MNRAS, 415, 2145
  \bibitem{2003A&A...403.1077K} K\"urster, M., Endl, M., Rouesnel, F., et al.  2003, A\&A, 403, 1077  
  \bibitem[Lenz  \& Breger(2004)]{Lenz04} Lenz, P., \& Breger, M.\ 2004,
  Period04: A Software Package to Extract Multiple Frequencies from Real Data,
  in {The A-Star Puzzle}, ed. J. Zverko, J. Ziznovsky, S.J. Adelman, \& W.W. Weiss,
   IAU Symp. 224 (Cambridge University Press),  786
  \bibitem{2009A&A...493..639M} Mayor, M., Udry, S., Lovis, C., et al. 2009, A\&A, 507, 487
  \bibitem{2011ApJ...726L...8P} Pierrehumbert, R. T. 2011, ApJ, 726, 8
  \bibitem{2012CeMDA.113...49T} Tadeu dos Santos, M., Silva, G.G., Ferraz-Mello,
  S., \& Michtchenko, T.A. 2012, Celestial Mechanics and Dynamical Astronomy, 113, 49
  \bibitem{2011A&A...528L...5T} Tuomi, M. 2011, A\&A, 528, 5
  \bibitem{2010ApJ...723..954V} Vogt, S.S., Butler, R.P., Rivera, E.J., et al.
          2010, ApJ, 723, 954 (V2010)
  \bibitem{2012AN....333..561V} Vogt, S.S., Butler, R.P., \& Haghighipour, N.
  2012, Astron. Nachr.,   333, 561 (VBH2012)
  \bibitem{2011A&A...528A.133V} von   Bloh, W., Cuntz, M., Frank, S., \& Bounama, C. 2011, A\&A, 528, A133
\end{thebibliography}
\end{document}